\documentstyle[sprocl,epsf]{article}

\newsavebox{\LSIM}
\sbox{\LSIM}{\raisebox{-1ex}{$\ \stackrel{\textstyle<}{\sim}\ $}}
\newcommand{\lsim}{\usebox{\LSIM}}
\newsavebox{\GSIM}
\sbox{\GSIM}{\raisebox{-1ex}{$\ \stackrel{\textstyle>}{\sim}\ $}}
\newcommand{\gsim}{\usebox{\GSIM}}

\def\be{\begin{equation}}
\def\ee{\end{equation}}
\def\bea{\begin{eqnarray}}
\def\eea{\end{eqnarray}}

\begin{document}

\title{SINGLETS AND THE ELECTROWEAK PHASE TRANSITION}

\author{S. J. HUBER\footnote{Presented at SEWM '98, Copenhagen, 2.-5.12.1998. 
Collaboration with M.~G.~Schmidt.}}

\address{Institut f\"ur Theoretische Physik, Universit\"at Heidelberg, \\
69120 Heidelberg, Germany\\E-mail: huber@thphys.uni-heidelberg.de} 




\maketitle\abstracts{Singlet extensions of the Standard Model (SM) allow for a
strongly first order electroweak phase transition, because of trilinear terms
in the tree-level potential. We present a systematic procedure 
to study the parameter space of the Next-to-Minimal Supersymmetric SM (NMSSM).
We find that this model is consistent with electroweak baryogenesis for a
wide range of parameters, allowing Higgs masses up to at least 115 GeV.}

\section{Introduction}
The order and strength of the electroweak phase transition  are
central questions in electroweak baryogenesis. Only in case of a
first order phase transition (PT) the associated departure from equilibrium 
is sufficient to induce  a relevant baryon number production. At the
critical temperature $T_c$ of a first order phase transition there 
exist two energetically degenerate phases which are separated by an 
energy barrier. In case of the electroweak phase transition (EWPT) these
phases differ in the vacuum expectation value (vev) of the Higgs
field $<h>$ and are referred to as ``symmetric'' ( $<h>=0$ ) and  
``Higgs'' phase ($<h>\neq0$). In the effective potential 
description this behavior is translated into two degenerate
minima which are separated by a bump. 

At a particular temperature  below $T_c$ Higgs phase bubbles, 
just large enough to grow (``critical bubbles''), 
nucleate and expand. After the
completion of the phase transition they fill all of space. As a bubble 
wall passes a point in space, the Higgs field changes rapidly which 
leads to a significant departure from equilibrium making it
possible to satisfy Sakharov's criteria.     

A further condition has to be satisfied: 
The anomalous baryon number violating processes which are 
essential for the baryon production during the phase transition may 
wash out the baryon asymmetry afterwards if they
are too weakly suppressed in the emerging Higgs phase at $T_c$. 
Since their rate is determined by the sphaleron energy which
is proportional to the Higgs vev, the washout criterion can be 
translated into 
$\frac{<h>(T_c)}{T_c} \gsim 1$.

Intensive studies of the electroweak phase transition in the past few 
years however 
showed that in case of the Standard Model (SM) this necessary
condition cannot be fulfilled. (For Higgs masses  compatible 
with experimental bounds there is no phase transition at all.) Successful
electroweak baryogenesis therefore requires extensions of the
minimal particle content 
in order to enlarge the cubic term in the effective potential that
triggers the first order phase transition.  
Very promising in this respect 
are models with additional scalar gauge singlet fields. 
Trilinear terms in the tree-level potential which arise
due to singlet-Higgs couplings 
should significantly strengthen the electroweak phase transition \cite{piet},
as long as the singlet and the Higgs are on the same (electroweak) scale.    

We have investigated the supersymmetric Standard Model with an additional
gauge singlet superfield $S$ (NMSSM). Its Higgs sector is characterized 
by the superpotential \cite{hunter,frog}
\begin{equation}W=\mu H_1H_2 +\lambda SH_1H_2-\frac{k}{3}S^3-rS \end{equation}
and the soft SUSY breaking terms 
\begin{eqnarray}
V_{\rm soft}&=&(BH_1H_2+{\lambda A_{\lambda}SH_1H_2-
\frac{k}{3}A_kS^3} +h.c.) \nonumber\\   
&&+m^2_{H_1}|H_1|^2+m^2_{H_2}|H_2|^2+m^2_S|S|^2 \ . 
\end{eqnarray}
Most interesting with respect to the strength of the phase transition
are the trilinear couplings $A_{\lambda}$ and $A_k$ since they 
contribute to the energy barrier that separates the symmetric and 
the Higgs phase. 

The well known domain wall problem is avoided by including the 
mass parameters $\mu,r,B$ that explicitly break the 
dangerous $Z_3$-symmetry. Furthermore, it was shown  
that in the $Z_3$-symmetric limit a viable phenomenology requires
couplings $\lambda,k \ll 1$ and  a 
singlet vev $<S>$ much larger than the Higgs vev \cite{ell}. 
In such a scenario the electroweak phase transition
is basically not modified by the presence of the singlet
and proceeds in the SM (MSSM)  way.
Therefore, strengthening the phase transition compared to the SM case
requires violation of the $Z_3$-symmetry.
Unfortunately, the general superpotential (1) leaves the $\mu$-problem,
it was originally designed for,
unsolved and there remains the question of destabilizing singlet tadpole 
divergences.

\section{ Renormalization Group Analysis}

We require the model to remain perturbative up to the GUT scale 
$M_{\rm GUT}$, where the  soft parameters are assumed to 
be characterized by a common gaugino mass $M_0$, a
universal trilinear coupling $A_0$ and a universal scalar mass
squared  $m_0^2$. The parameters at different
scales are related via the renormalization group equations (RGEs),
which we approximate to 1-loop order. 

\begin{figure}[t] 
\begin{picture}(0,100)
\put(-80,-290){\epsfxsize12cm \epsffile{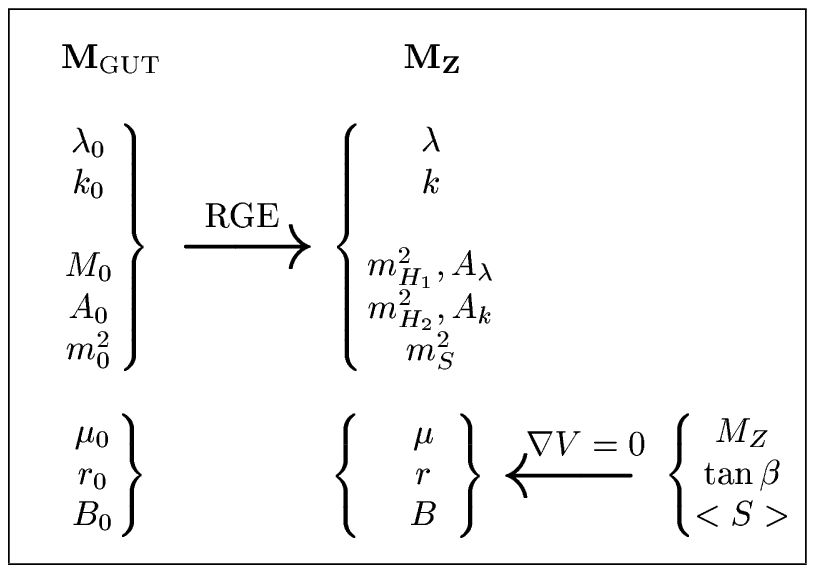}}
\put(95,-185){\epsfxsize10cm \epsffile{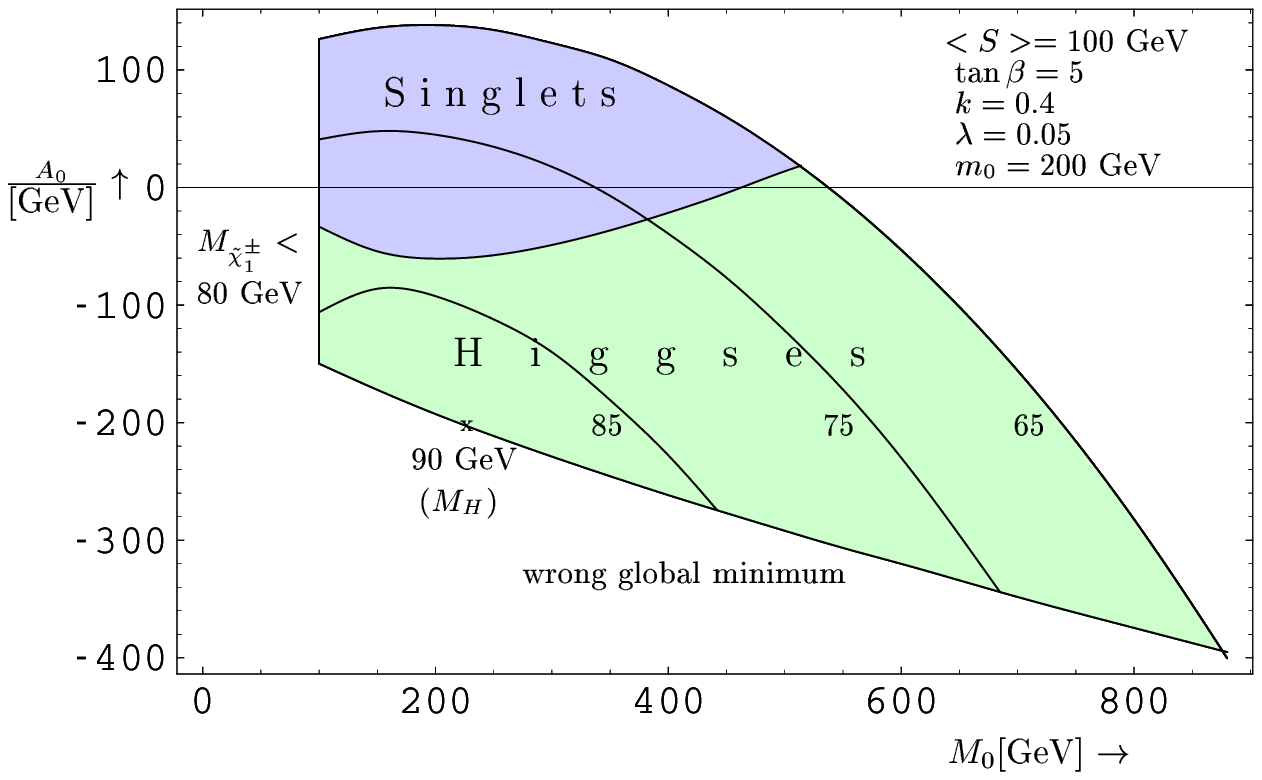}}
\put(65,104){\small (a)}
\put(250,103){\small (b)}
\end{picture} 
\caption{(a): Sketch of our procedure to determine the
weak scale as function of the GUT scale parameters. (b):
Constrained parameter range in the $M_0$-$A_0$ plane, while
the remaining parameters are fixed.}
\label{fig3} 
\end{figure}

Instead of simply studying the phenomenology of a 
randomly chosen GUT scale parameter set (``random shooting''),
which is a rather inefficient method because there usually appear some light
unobserved SUSY particles in the spectrum, we use the more
elaborated procedure sketched in fig. 1a \cite{hub}. 
Our method determining the parameters at the weak scale 
which combines RGEs and extremal conditions for the 
effective potential
$V(H_1,H_2,S)$ is based 
on the decoupling of
$\mu,r,B$ from the RGEs of the other parameters. Therefore, after
fixing the values of $\lambda_0,k_0,M_0,A_0,m_0^2$ we can calculate
all parameters in $V(H_1,H_2,S)$ with exception of $\mu,B,r$. These
we determine via the extremal condition $\nabla V(H_1,H_2,S)=0$ postulated 
at some minimum characterized by $(M_Z,\tan\beta,<S>)$. 
For the elimination of $\mu,r,B$ we use the  1-loop Higgs potential
$V(H_1,H_2,S)$ at zero temperature where the contributions of tops,
stops and gauge bosons are taken into account. 
Our procedure  allows for an efficient and systematic study of the 
remaining seven dimensional parameter space 
\[\tan\beta,<S>,\lambda_0,k_0,M_0,A_0,m_0^2.\]

In the following we perform cuts in the
$M_0$--$A_0$ plane since these parameters determine the trilinear
couplings $A_{\lambda}$, $A_k$ which  regulate the strength of the 
electroweak phase transition in this model.
The phenomenologically viable parameter space is
characterized by the following conditions:
\begin{itemize}
\item The extremum used in the elimination procedure discussed above 
has to be the global minimum of the Higgs potential which results in
   the lower bound on $A_0$ in fig. 1b.
\item There should exist no light unobserved particles, especially
   the mass of the lightest chargino\footnote{Our choice of $m^2_0$ 
prevents the appearance of any light sfermions.} 
has to obey  $M_{\tilde{\chi}^{\pm}_1}>80$ GeV corresponding to the
 lower bound on $M_0>100$ GeV in fig. 1b.
\end{itemize} 
Of particular interest are the properties of the lightest
neutral CP-even 'Higgs' boson  mass eigenstate $H$, which is a
mixing of the Higgses and the singlet.
Its mass $M_H$ is maximized for low values of $A_0$. 
From fig. 1b one can also read off the parameter region 
where the lightest 'Higgs' is predominantly a singlet.
In this case the experimental bound on $M_H$ is certainly lower 
than the 95 GeV for Standard Model like Higgses, therefore we 
also kept smaller Higgs masses down to 65 GeV.

\begin{figure}[t] 
\begin{picture}(0,100)
\put(-78,-183){\epsfxsize10cm \epsffile{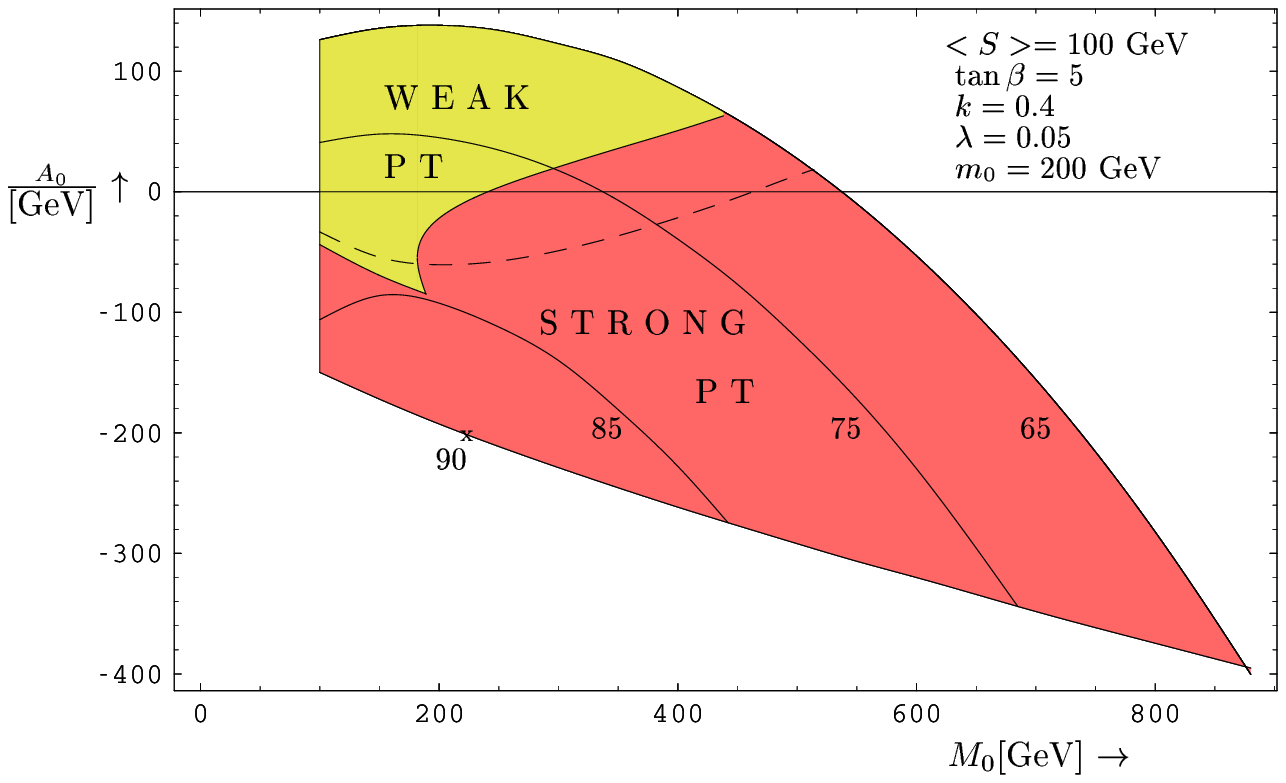}}
\put(97,-220){\epsfxsize10cm \epsffile{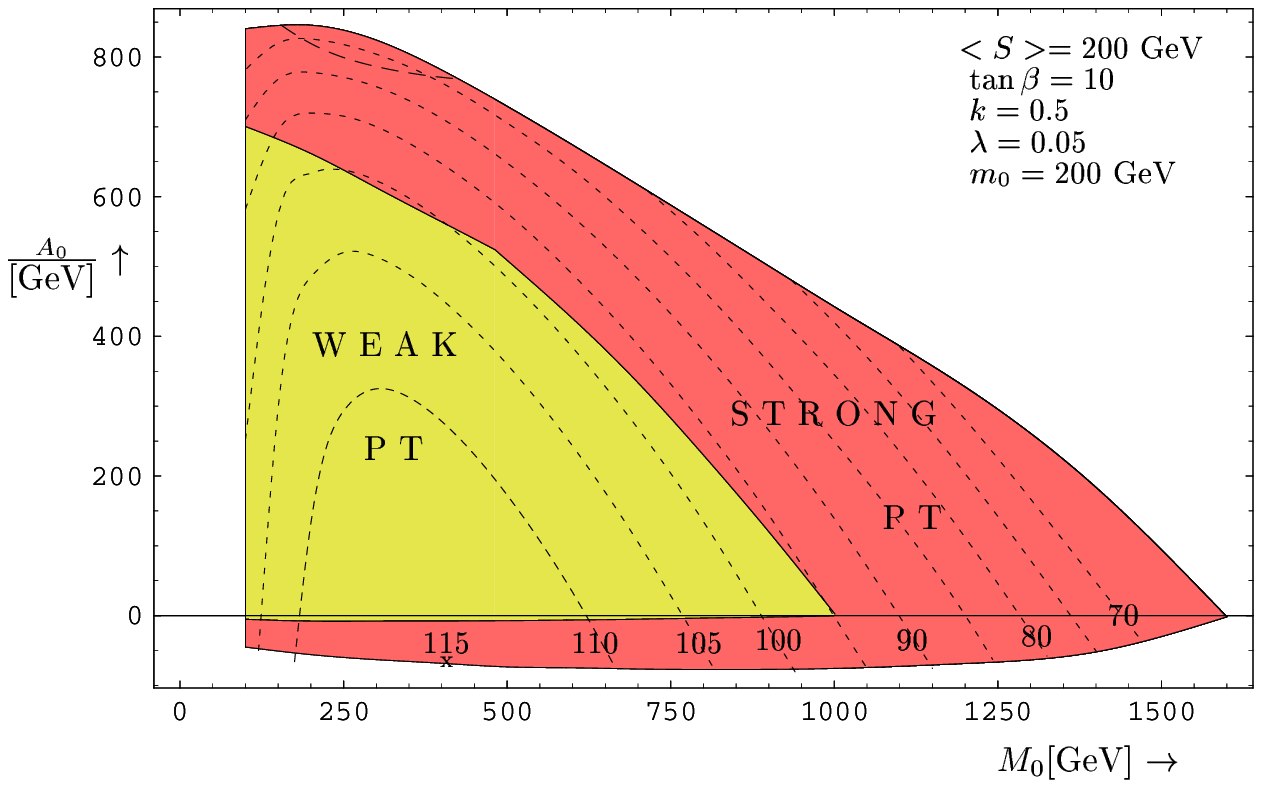}}
\put(85,104){\small (a)}
\put(260,103){\small (b)}
\end{picture} 
\caption{Strength of the EWPT as a function of $M_0,A_0$ for two 
different sets of fixed parameters ($<S>=100,200$ GeV).}
\label{fig2} 
\end{figure}

\section{The Electroweak Phase Transition}
In order to examine the strength of the phase transition, one
has to take into account thermal corrections to the
effective Higgs potential $V(H_1,H_2,S)$. We include the 1-loop contributions of 
tops, stops, gauge bosons, Higgs bosons, neutralinos and charginos.
Since we can rely on the tree-level cubic terms, the most important
effect of finite temperature is the appearance of  effective thermal 
masses \cite{piet,frog}:
\[m^2 \longrightarrow m^2+\mbox{const}\cdot T^2\]
From the thermal effective potential $V_T(H_1,H_2,S)$ we determine 
the  critical temperature $T_c$ where there exist 
two degenerate minima, a symmetric one with $<H_i>=0$ and
a broken minimum with $<H_i>\equiv v_{i,c}\neq 0$.  With this information
we can check the washout criterion for electroweak baryogenesis: 
\begin{equation}\frac{\sqrt{v_{1,c}^2+v_{2,c}^2}}{T_c}\gsim 1 \quad . \end{equation}

Fig. 2a displays the strength of the EWPT for the parameter set already 
presented in fig. 1b \cite{hub}. We find that in a large part of the 
parameter space  the washout condition is satisfied (``strong PT''),
while Higgs masses up to 90 GeV are accessible. In fig. 2b we have 
increased $<S>,\tan\beta$ in order to obtain larger values of 
$M_H\lsim 115$ GeV. We observe that the parameter region compatible
with electroweak baryogenesis (3) becomes smaller in this case.
Notice, that both parameter sets used in fig. 2 satisfy  
$<S>\sim {\cal O}(M_Z)$, otherwise we would just obtain
the SM phase transition.

\section{Summary and Outlook}
We presented a method to systematically study  the NMSSM parameter space,
avoiding random shooting. Our investigation of the EWPT
shows that a considerable range of parameters exhibits a phase transition
strong enough for electroweak baryogenesis, where 
Higgs masses up to at least 115 GeV are allowed. 

Having found a parameter set with a strongly first order EWPT, one can start
calculating the arising baryon asymmetry. To tackle this problem one has to 
address the questions of CP-violation (talk of A.~Davies, this conference)
and the dynamics of the nucleating bubbles (poster of P. John, this conference).

\section*{Acknowledgments}
I would like to thank M.~G.~Schmidt for enjoyable collaboration
and also P.~John for a lot of useful discussions. This work was supported
in part by the TMR network {\em Finite Temperature Phase Transitions
in Particle Physics}, EU contract no. ERBFMRXCT97-0122.

\end{document}